# Multipath Interference Suppression in Indirect Time-of-Flight Imaging via a Novel Compressed Sensing Framework

Yansong Du[1], YuTong Deng[1], Yuting Zhou[1], Feiyu Jiao[1], Bangyao Wang[1], Zhancong Xu[1], Zhaoxiang Jiang[2,3],　And Xun Guan[1,4]

1Tsinghua Shenzhen International Graduate School, Tsinghua University, Shenzhen, 581055, China

2Guangdong Laboratory of Artificial Intelligence and Digital Economy (SZ), Shenzhen, 518060, China

3jiangzhaoxiang@gml.ac.cn

4xun.guan@sz.tsinghua.edu.cn

**Abstract:** We propose a novel compressed sensing method to improve the depth reconstruction accuracy and multi-target separation capability of indirect Time-of-Flight (iToF) systems. Unlike traditional approaches that rely on hardware modifications, complex modulation, or cumbersome data-driven reconstruction, our method operates with a single modulation frequency and constructs the sensing matrix using multiple phase shifts and narrow-duty-cycle continuous waves. During matrix construction, we further account for pixel-wise range variation caused by lens distortion, making the sensing matrix better aligned with actual modulation response characteristics. To enhance sparse recovery, we apply K-Means clustering to the distance response dictionary and constrain atom selection within each cluster during the OMP process, which effectively reduces the search space and improves solution stability. Experimental results demonstrate that the proposed method outperforms traditional approaches in both reconstruction accuracy and robustness, without requiring any additional hardware changes.

## 1. Introduction

Time-of-Flight (ToF) cameras have become a cornerstone in modern 3D imaging applications due to their ability to capture dense depth information with low cost, compact form factor, and real-time performance[1]. In particular, indirect ToF (iToF) systems, typically employing amplitude-modulated continuous wave (AMCW) modulation, estimate depth by measuring the phase shift between emitted and received light signals[2]. This architecture enables efficient and high-resolution depth sensing and is widely adopted in robotics, augmented reality, and autonomous systems[3]. Despite these advantages, iToF cameras suffer from several limitations, most notably environment-induced non-systematic errors[4]. Among them, multi-path interference (MPI) is a primary source of error, arising when reflected light from multiple surfaces reaches the sensor simultaneously, causing ambiguous or biased phase measurements. These errors are especially pronounced in environments with highly reflective or translucent materials, leading to significant degradation in depth accuracy[5-9].

　　To address the above challenges, prior research has mainly focused on two directions: hardware-oriented enhancements and data-driven approaches. Some studies aim to improve system performance by modifying pixel sensor structures [10], increasing modulation frequencies, or optimizing analog front-end circuits [11]. While these methods perform well in controlled environments, they often introduce high system complexity and engineering cost, which limits their practicality. In parallel, data-driven methods rely on deep learning models trained on large annotated datasets to correct distorted depth information [12–14]. Although these models perform well in familiar scenarios, they are highly sensitive to the data distribution and tend to generalize poorly to unseen scenes or distribution shifts. While both directions have achieved promising results under specific conditions, their reliance on expensive hardware or large-scale annotations still hinders their broader deployment in real-world applications [15–17].

Against this backdrop, Compressed Sensing (CS) has emerged as a promising theoretical framework that balances system simplicity and robustness. Compared with the above approaches, CS leverages sparse modeling and optimization-based recovery to achieve efficient signal reconstruction from limited measurements, offering strong anti-interference capabilities and theoretical interpretability. As such, it shows great potential for depth sensing in complex or dynamic environments. However, existing CS methods still face several challenges. On one hand, sensing matrix design often struggles to balance sensing performance and physical realizability. While structured matrices improve deployability, they may introduce high mutual coherence and limited representation power [18-24]. On the other hand, mainstream sparse recovery algorithms typically adopt pixel-wise modeling, ignoring spatial structures and contextual dependencies, and tend to become unstable under high-interference conditions [25-28]. Moreover, recent approaches that incorporate multi-frequency modulation or segmented imaging to enhance multi-target separation [29-31] often involve complex modulation control and post-processing pipelines, requiring manual parameter tuning and limiting applicability in low-cost, automated systems.

In this work, we propose a compressed sensing-based framework to enhance depth reconstruction accuracy and multi-target separation in indirect Time-of-Flight systems affected by multi-path interference. Our method operates with a single modulation frequency, avoiding the complexity of multi-frequency schemes commonly adopted in prior work. We construct the sensing matrix using multiple phase shifts of narrow-duty-cycle continuous waves, and importantly, we account for pixel-wise range variation caused by lens distortion, enabling the sensing model to more accurately reflect real modulation responses across the sensor. To improve sparse recovery, we introduce a clustering-constrained reconstruction strategy: the distance response dictionary is divided using K-Means clustering [32], and the Orthogonal Matching Pursuit (OMP) algorithm [33] is constrained to select atoms only within each cluster. This significantly reduces the search space and improves reconstruction stability. The proposed approach requires no additional hardware modifications and can be seamlessly integrated into existing iToF systems, offering strong practicality and deployment flexibility.

## 2. Principles and analysis

As shown in Figure 1, in a typical indirect Time-of-Flight (iToF) imaging system, a modulated light source emits a periodic modulation signal s(ϕ), which propagates through the scene and is partially reflected by various surfaces, resulting in a returned signal g(ϕ) received by the sensor. To estimate depth, the system configures a set of reference phase shifts $\phi_t$, and for each phase shift, it performs correlation operations between the received signal and the reference modulation signal. This process yields a set of phase-shifted measurements, which encode the temporal delay of the reflected light. Mathematically, this imaging process can be described by:

$$c(\phi_t) = \int_0^{2\pi} s(\phi) \cdot g(\phi + \phi_t) d\phi \tag{1}$$

In real-world scenarios, however, the presence of reflective surfaces such as walls, glass, or other scattering structures introduces multi-path interference (MPI), which severely degrades depth estimation accuracy. Under MPI, the reflected signal is no longer a simple copy of the main path return, but rather a superposition of the primary (direct) path signal and multiple secondary (indirect) path signals. This can be modeled as:

$$s(\phi) = s_{main}(\phi) + \sum s_{noise}(\phi) \tag{2}$$

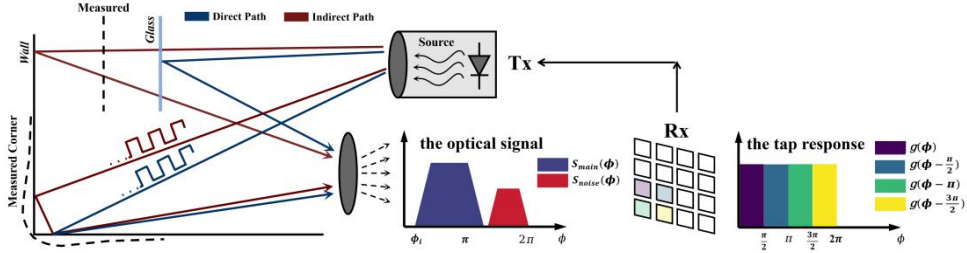

Fig.1: Illustration of multi-path interference in indirect Time-of-Flight (iToF) ranging

In typical indirect Time-of-Flight (iToF) imaging systems, when narrow-duty-cycle modulation is employed, the modulation waveform s(ϕ) can be approximated as a sparse pulse sequence. Under such conditions, the entire imaging process can be reformulated as a convolutional sampling procedure. This allows the measurement model to be expressed within the compressed sensing framework as a linear observation process:

$$c = A \cdot g \tag{3}$$

Here, $c \in C^{L \times 1}$ denotes the complex-valued observation vector acquired under *L* reference phases $\{\phi_1, \phi_2, ..., \phi_m\}$, $g \in C^{M \times 1}$ represents the sparse back-scattered intensity distribution to be estimated. The Measurement Matrix $A \in C^{L \times M}$ is the Measurement Matrix, constructed as follows:

$$A = \begin{bmatrix} e^{j(2\pi f \tau_1 - \phi_1)} & \cdots & e^{j(2\pi f \tau_L - \phi_1)} \\ e^{j(2\pi f \tau_1 - \phi_2)} & \cdots & e^{j(2\pi f \tau_L - \phi_2)} \\ \vdots & \ddots & \vdots \\ e^{j(2\pi f \tau_1 - \phi_M)} & \cdots & e^{j(2\pi f \tau_L - \phi_M)} \end{bmatrix} \qquad (4)$$

The construction of this matrix follows the phase modulation principle in iToF imaging. Specifically, under a fixed modulation frequency, each column corresponds to a distance bin $\tau_L$, representing its modulation response across all reference phases. Each row corresponds to a fixed reference phase $\phi_m$, reflecting the contributions from different distance bins. Overall, the matrix encodes the phase response characteristics of depth-resolved reflections, illustrating how the system performs depth sampling across the scene.

In addition, we incorporate a disturbance term to account for residual uncertainties and system mismatches. The final measurement model becomes:

$$c = A \cdot g + \varepsilon \qquad (6)$$

The term $\varepsilon$ denotes structural perturbations in the observations, which reflect discrepancies between actual modulation responses and the ideal model—particularly pronounced near depth discontinuities due to severe phase deviation. While such disturbances may potentially reduce the column-wise orthogonality of $A$, they also introduce greater variability among columns. This increased diversity is beneficial for satisfying the Restricted Isometry Property (RIP), a key condition in compressed sensing theory, thereby enhancing the robustness and reliability of sparse reconstruction in practical iToF imaging scenarios.

Under the sparse modeling framework proposed in this work, the observed response is reconstructed into a discrete multipath response vector, where the horizontal axis corresponds to different depth values, and the vertical axis represents the signal intensity along each path. This sparse structure enables the separation of the primary reflection and multipath interference along the depth dimension, thereby improving phase demodulation accuracy and laying a solid foundation for subsequent depth denoising and artifact suppression.

To validate the effectiveness of the proposed model under multipath interference conditions, we simulated the mixed measurements acquired by an iToF camera, which include both the primary reflection and multiple interference paths, as shown in the left panel of Fig. 2. Based on the constructed measurement matrix and the compressed sensing algorithm, we then performed sparse recovery of the overlapped signal and reconstructed the complete multipath response structure. As illustrated in the right panel of Fig. 2, the primary and interference signals are accurately separated along the depth dimension, and the reconstructed result closely matches the ground-truth signal in terms of depth locations. Error analysis shows a reconstruction error of 3.36%, further demonstrating the robustness and accuracy of the proposed method under complex path conditions.

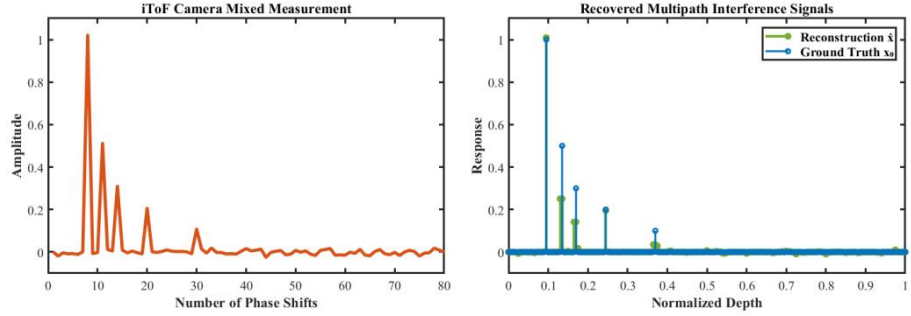

Fig .2: Reconstruction of Multipath Interference Signals from Mixed iToF Measurements via Sparse Modeling

## 3. Calibration of the Measurement Matrix

We validated the proposed compressed sensing framework for enhancing depth reconstruction and multi-target separation in indirect Time-of-Flight (iToF) systems through a systematic experimental procedure, beginning with the calibration of the measurement matrix. This matrix establishes the mapping between modulation patterns and received signals, serving as the foundation for subsequent phase retrieval and sparse reconstruction.

At the initial stage of calibration, we configured the light source to output a modulated waveform with a 5% duty cycle, forming a narrow-duty-cycle modulated continuous wave. This signal exhibits a pulse-like shape in the time domain, characterized by a short on-time and steep rising and falling edges, thereby enhancing phase sensitivity along the depth axis. As shown in Fig. 3(a), the waveform exhibits favorable phase modulation characteristics.

(a)            (b)            (c)

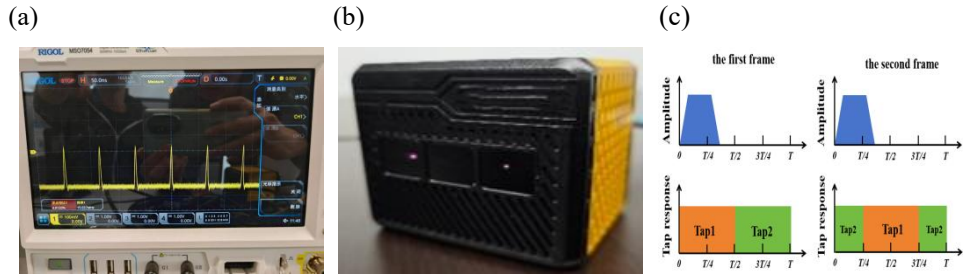

Fig .3: (a) Modulated continuous light waveform with narrow duty cycle. (b) The custom-built iToF camera prototype used in experiments. (c) The working principle of the Sony IMX570 indirect ToF sensor.

The camera's exposure time was fixed throughout the entire experiment in order to eliminate potential signal variations caused by automatic exposure adjustments. The iToF camera used in this work, as shown in Fig. 3(b), was independently developed by our research group and is equipped with a Sony IMX570 CMOS image sensor[34]. This sensor adopts a dual-tap pixel architecture, allowing it to capture the $0°/180°$ and $90°/270°$ phase pairs over two consecutive frames. By emulating a four-tap demodulation scheme, this design significantly enhances the accuracy of phase-based depth estimation. The modulated illumination is provided by a VCSEL (Vertical-Cavity Surface-Emitting Laser) light source, which is driven by a CXD4029 modulation driver. The internal working principle of the IMX570 sensor, including its dual-frame phase acquisition mechanism, is illustrated in Fig. 3(c).

We performed spatial scanning by moving the target object from 300 mm to 1300 mm in 1 mm increments. At each fixed distance, the iToF camera adjusted the delay between the transmitter (TX) and receiver (RX) to acquire 20 sets of responses under different phase shift configurations. For each configuration, four correlation images corresponding to $0°$, $90°$, $180°$, and $270°$ were captured, introducing rich phase modulation diversity without altering the object distance. This mechanism enables the generation of more generalizable observation data at each position, thereby improving the coverage and expressiveness of the modulation-to-response mapping. As illustrated in Fig. 4, this process constructs the measurement matrix corresponding to the center pixel, which can subsequently be extended to other pixels based on the camera's distortion model to form a complete set of measurement matrices.

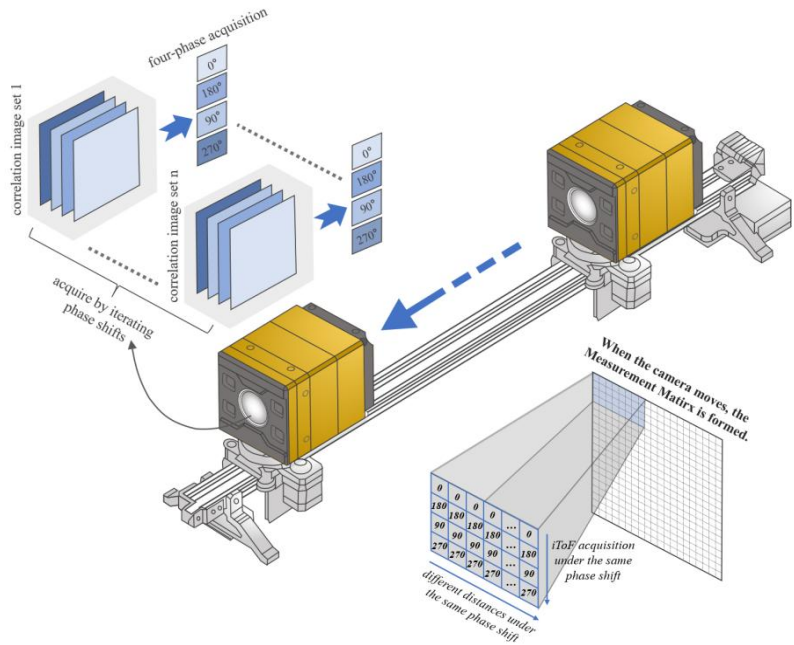

**Fig .4: Experimental procedure for measurement matrix calibration via spatial scanning and phase modulation diversity in an iToF system**

However, in practical systems, due to lens distortion, the incident angles corresponding to central and peripheral pixels differ significantly. This results in deviations between the actual modulation responses and the idealized model, introducing structural inconsistencies in the measurement matrix and compromising the accuracy and stability of multipath signal reconstruction and depth estimation.

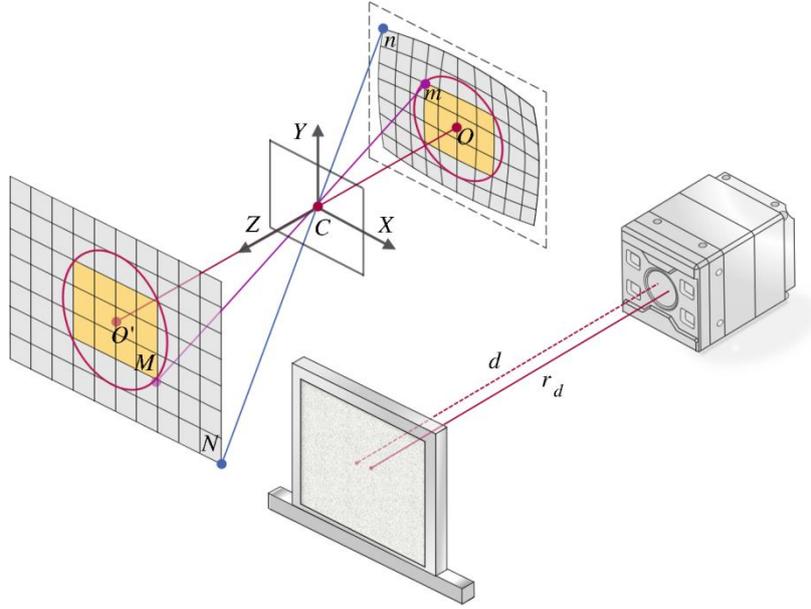

**Fig .5: Geometric interpretation of depth errors caused by pixel-wise distortion in an iToF imaging system**

We propose a per-pixel correction mechanism based on true physical depth to address this issue. Specifically, each pixel coordinate $(u,v)$ is projected onto a normalized angular coordinate system $(x,y)$, and the angle of deviation from the optical axis is computed using the intrinsic camera parameters. Based on this deviation, we apply a correction model to adjust the propagation path length:

$$r_d(u,v,d) = d \cdot \sqrt{1+\Delta(x,y)} \qquad (5)$$

Here, $\Delta(x,y)$ represents a path elongation factor induced by the angular deviation, accounting for the nonlinear propagation distortion caused by oblique incidence. This correction function enables consistent modulation responses across pixels for the same physical depth, effectively compensating for phase errors caused by varying view angles. The correction mechanism is then integrated into the construction of the measurement matrix $A$, as illustrated in Fig. 5, ensuring that each pixel's modulation response aligns more closely with its actual geometric propagation path. This leads to improved spatial consistency in the sensing model and enhances the overall accuracy of depth recovery under real-world optical conditions.

Under each phase shift configuration, the system continuously captured 100 frames using consistent acquisition parameters and performed frame averaging to enhance measurement stability and repeatability. Meanwhile, baseline images were acquired under TX-off conditions to eliminate background interference such as dark current, ambient light, and fixed-pattern noise. By performing frame-wise subtraction between the modulated and baseline images, we extracted the effective responses purely driven by the modulated signal, which were used to construct the calibration Measurement Matrix. To further verify the consistency of illumination conditions and sensor responses during calibration, representative pixels from the center and periphery of the sensor were selected, and their four-phase responses were analyzed under multiple distance settings. As shown in Fig. 6, the left panel illustrates the four-phase intensity responses of the central and peripheral pixels under a specific phase shift configuration, reflecting spatial consistency in their responses. The right panel presents the intensity ratios of peripheral pixels relative to the central pixel across different distances and computes their average errors. Experimental results demonstrate that the measured responses exhibit minimal fluctuation with distance variation, and the average intensity comparison error among different pixels remains below 5%, confirming the system's high stability in both illumination and sensor response—thus providing a solid foundation for constructing an accurate and robust Measurement Matrix.

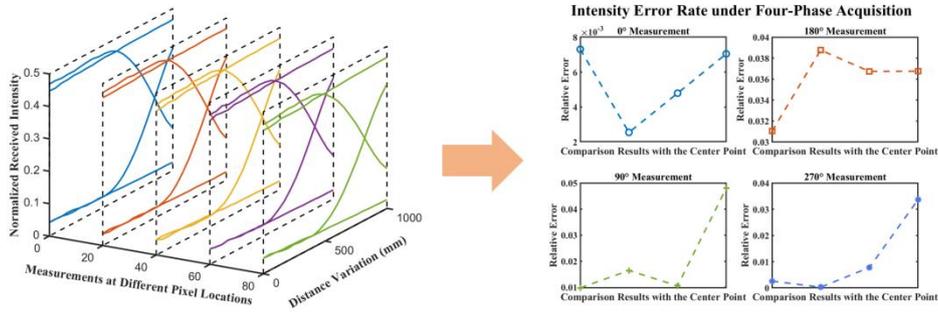

**Fig .6: Evaluation of Four-Phase Response Consistency Across Spatial Pixel Locations During Calibration**

Upon completion of the calibration process, the measurement matrix was employed to decode the received modulated signals and extract the depth response vector for each pixel. Leveraging sparse signal recovery techniques, the multipath interference components were effectively suppressed based on their amplitude characteristics, allowing the accurate preservation of the valid depth information associated with the primary reflection path.

## 4. Sparse Reconstruction Algorithm

After system calibration, we employ a structure-aware sparse reconstruction strategy to extract stable and accurate depth information from modulation responses affected by multipath interference. The overall process is illustrated in Fig. 7. Let the observed response vector at a given pixel be denoted as $c \in R^M$, and the corresponding measurement matrix as $A \in R^{M \times N}$, where each column $a_j$ represents the modulation response associated with a specific depth bin. Due to spatial redundancy among modulation responses, conventional sparse reconstruction methods may suffer from ambiguity and instability when the atoms in A are highly correlated.

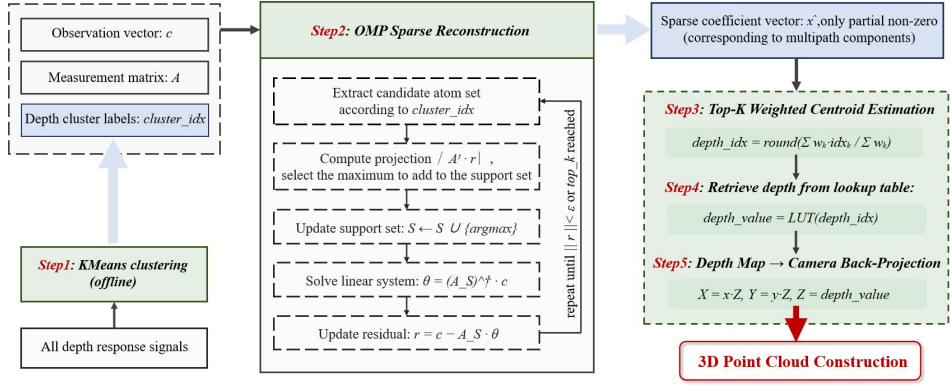

**Fig .7: Flowchart of Sparse Signal Recovery and Multipath Interference Removal Algorithm**

We mitigate this issue by introducing a clustering-guided atom selection mechanism, in which K-Means clustering is applied to the dictionary . Atoms within the same cluster exhibit similar response patterns and depth properties. This cluster label, obtained during preprocessing, is used to constrain the candidate atoms during sparse coding for each pixel, thereby improving the robustness and discriminability of the reconstruction process. During sparse coding, we apply the Orthogonal Matching Pursuit (OMP) algorithm within the sub-dictionary $A_c \subseteq A$ corresponding to the pixel's cluster. At each iteration k, the atom $a_j$ that exhibits the highest correlation with the current residual $r^{(k)}$ is selected:

$$j^* = \arg\max_{a_j \in A_{c^*}} \left| \langle a_j, r^{(k)} \rangle \right| \tag{6}$$

And added to the support set S. Sparse coefficients are then updated via least squares:

$$x_S = \arg\min_{x} \left\| A_S x - c \right\|_2^2 \tag{7}$$

with residuals updated as:

$$r^{(k+1)} = c - A_S x_S \tag{8}$$

After sparse decoding, the nonzero coefficients correspond to multiple potential modulation response components. Given that neighboring atoms often yield similar magnitudes—making it difficult to isolate a single target—we first extract the Top-K components with the largest magnitudes and local proximity in index space. We then compute a weighted centroid based on their coefficients to estimate a continuous depth index:

$$depth\_idx = \frac{\sum_{j \in K} x_j \cdot j}{\sum_{j \in K} x_j} \tag{9}$$

Where K denotes the index set of selected atoms and $x_j$ is the corresponding coefficient magnitude.

The estimated depth index is further mapped to a real-valued depth using a lookup table. Finally, the depth map is back-projected into 3D space using the pinhole camera model with intrinsic parameters. This approach effectively separates the primary reflection path from multipath interference components while maintaining low reconstruction complexity. It is particularly well-suited for modulation-response-driven depth estimation tasks, demonstrating strong robustness and high accuracy in complex interference scenarios.

## 5. Result and Comparison

In order to evaluate the effectiveness of the proposed multipath interference suppression method under complex real-world conditions, we designed two representative experimental scenarios, as illustrated in Fig. 8. The first scenario involves a corner structure, which introduces strong reflections and cross-structure multipath propagation. The second scenario captures a wall through glass, representing the interference caused by transparent media and non-line-of-sight paths. Both settings are known to produce severe MPI artifacts in iToF systems, making them representative test cases. A checkerboard was used for geometric calibration to obtain high-precision ground truth 3D point clouds, which served as the reference. Meanwhile, unprocessed raw depth data were captured by an iToF camera for evaluating the reconstruction performance of different methods.

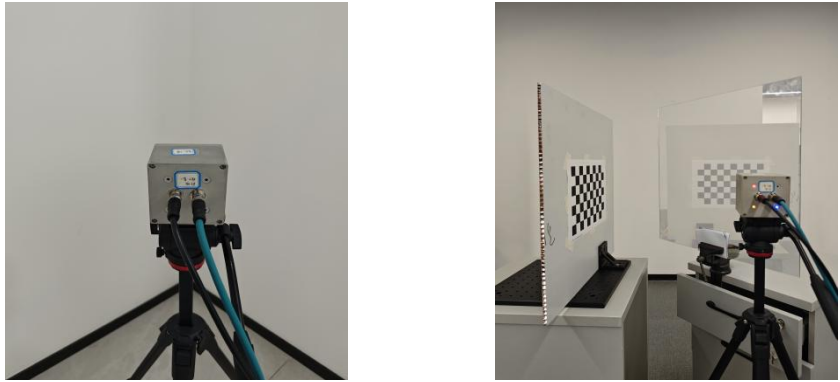

**Fig .8: Experimental Setup for Multipath Interference: (Left) Corner scenario; (Right) Glass reflection scenario**

The Fig .9 presents the visual comparison among the raw point cloud, our reconstruction result, and the ground truth. In the corner scenario, the raw point cloud exhibits significant curvature and non-physical bulging at the edge intersection due to MPI. In contrast, our method successfully restores the sharp and clear corner geometry, showing high consistency with the ground truth in terms of edge continuity and overall shape. In the glass scenario, the raw depth data suffer from large-scale fluctuations, indicating depth instability, whereas our method significantly improves the planar consistency, with the reconstructed surface closely matching the ground truth in both elevation and spatial distribution. The cross-sectional depth profiles further confirm this observation: the curve from our method closely aligns with the ground truth, while the raw depth deviates significantly.

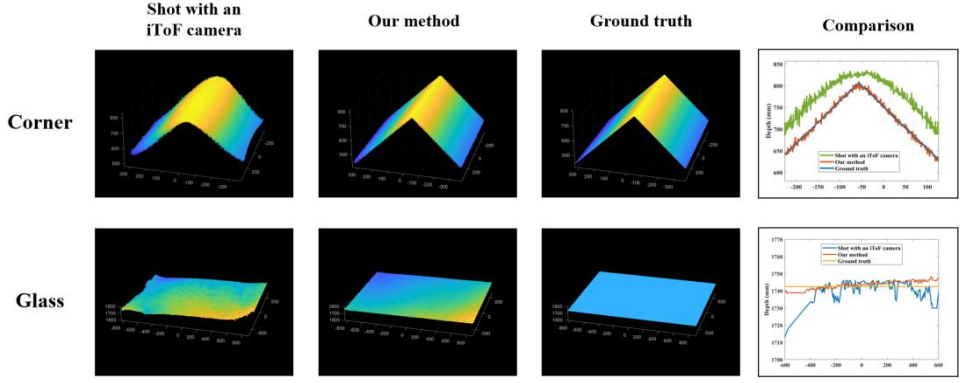

Fig .9: Visual and cross-sectional comparison of reconstructed depth in two multipath interference scenarios (Corner and Glass). From left to right: raw depth captured by iToF camera, reconstruction by our method, ground truth, and depth profile comparison.

A comprehensive quantitative comparison was further conducted with four representative methods, including two learning-based models (SHARP-Net[35] and ToFNet[36]) and two compressive sensing-based algorithms (CoSaMP[37] and FISTA[38]). To ensure a fair and rigorous evaluation, we adopted three widely used metrics: Mean Absolute Error (MAE), Root Mean Square Error (RMSE), and Structural Similarity Index (SSIM), which respectively measure local accuracy, global deviation, and structural consistency. These metrics are defined as follows:

·**Mean Absolute Error (MAE):**

$$MAE = \frac{1}{N}\sum_{i=1}^{N}\left|\hat{d}_i - d_i\right| \quad (10)$$

where $\hat{d}_i, d_i$ denote the predicted and ground truth depth values, and N is the number of valid pixels.

·**Root Mean Square Error (RMSE):**

$$RMSE = \sqrt{\frac{1}{N}\sum_{i=1}^{N}\left(\hat{d}_i - d_i\right)^2} \quad (11)$$

·**Structural Similarity Index (SSIM):**

$$SSIM = \frac{2\mu_{\hat{d}}\mu_d + c_1}{\mu_{\hat{d}}^2 + \mu_d^2 + c_1} \cdot \frac{2\sigma_{\hat{d}d} + c_2}{\sigma_{\hat{d}}^2 + \sigma_d^2 + c_2} \quad (12)$$

where $\mu_{\hat{d}}$ and $\mu_d$ are the local means, $\sigma_{\hat{d}}^2$ and $\sigma_d^2$ are the local variances, $\sigma_{\hat{d}d}$ is the local covariance between the reconstructed depth $\hat{d}$ and the ground truth $d$ and $c_1, c_2$ are constants used to avoid instability when the denominators are small.

The comparison is conducted across these methods in two typical multipath interference scenarios: a wall corner and a glass surface. As illustrated in Fig. 10, both qualitative and quantitative results consistently demonstrate that our method achieves the best performance in all evaluation metrics across both scenarios.

Specifically, in the Corner scene, our method obtains a Mean Absolute Error (MAE) of 3.76 mm, a Root Mean Square Error (RMSE) of 15.24 mm, and a Structural Similarity

Index (SSIM) of 0.94998, significantly outperforming other methods in both accuracy and structural fidelity. In the Glass scene, where diffuse and internal reflections often degrade performance, our method again achieves the lowest errors and highest similarity, with an MAE of 3.72 mm, RMSE of 10.64 mm, and SSIM of 0.9922. These results clearly indicate the robustness and generalization ability of our approach under challenging multipath conditions.

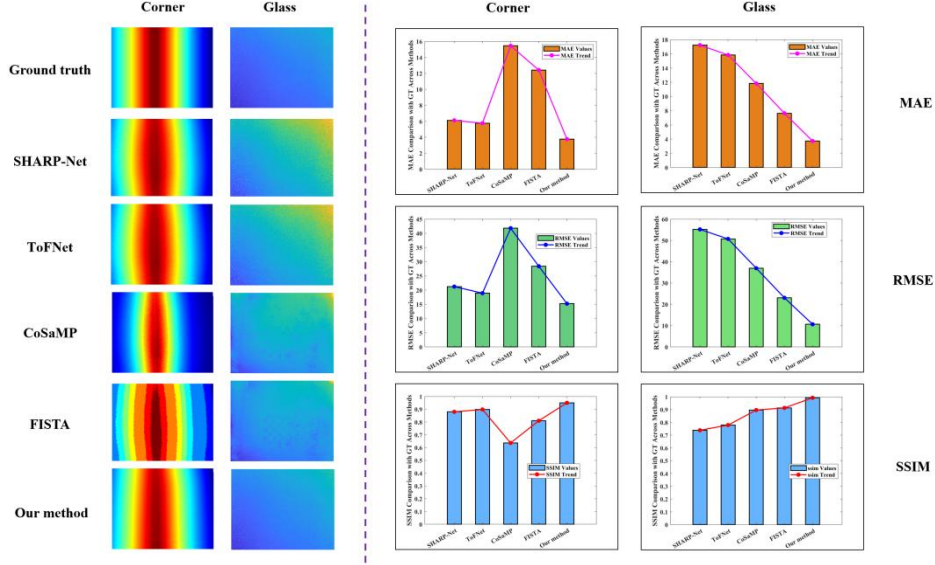

**Fig .10: Qualitative and quantitative comparison of different methods in two multipath interference scenarios (Corner and Glass). The left columns show reconstructed depth maps; the right column presents evaluation results using MAE, RMSE, and SSIM metrics.**

## 6 . Conclusion

Multipath interference (MPI) poses a major challenge to the depth accuracy of indirect Time-of-Flight (iToF) cameras due to its non-systematic nature. This paper presents a compressed sensing-based framework for effectively suppressing MPI in iToF imaging. By utilizing a single modulation frequency with multiple phase shifts and narrow-duty-cycle continuous waves, the method avoids the complexity of multi-frequency schemes while preserving measurement diversity. To further enhance reconstruction fidelity, pixel-level corrections are applied to account for lens distortion-induced range variations, ensuring better alignment between the sensing matrix and actual modulation responses. Additionally, K-Means clustering is introduced to guide atom selection in Orthogonal Matching Pursuit (OMP), significantly improving the stability and sparsity of the reconstruction process. Extensive experiments conducted on two representative MPI-prone scenarios—wall corner and glass surface—demonstrate that the proposed method achieves superior reconstruction accuracy across all evaluation metrics. In the corner scenario, the method attains an MAE of 3.76 mm, RMSE of 15.24 mm, and SSIM of 0.94998, while in the more challenging glass scenario, it further achieves an MAE of 3.72 mm, RMSE of 10.64 mm, and SSIM of 0.9922. These results consistently outperform both deep learning-based and traditional compressed sensing methods. Without requiring any hardware modifications, the proposed approach can be seamlessly integrated into existing iToF systems, offering strong adaptability and practical applicability in real-world environments.

**Funding.**
National Natural Science Foundation of China (U23A20282); Shenzhen Science, Technology and Innovation Commission (KJZD20231023094659002, JCYJ20220530142809022, WDZC20220811170401001).

**Disclosures.**
The authors declare no conflicts of interest.

**Data Availability.**
Data underlying the results presented in this paper are not publicly available at this time but may be obtained from the authors upon reasonable request

**Reference:**
[1] X. Su and Q. Zhang, "Dynamic 3-D shape measurement method: a review," Opt Lasers Eng 48(2), 191–204 (2010).
[2] S. Foix, G. Alenya, and C. Torras,"Lock-in Time-of-Flight (ToF) Cameras: A Survey," IEEE Sensors J. 11(9), 1917–1926 (2011).
[3] A. Bhandari, C. Barsi, and R. Raskar, "Blind and reference-free fluorescence lifetime estimation via consumer time-of-flight sensors,"Optica 2(11), 965 (2015).
[4] Yansong Du, Zhaoxiang Jiang, Jindong Tian, and Xun Guan, "Modeling, analysis, and optimization of random error in indirect time-of-flight camera," Opt. Express 33, 1983-1994 (2025)
[5] Agresti, Gianluca, and Pietro Zanuttigh. "Combination of spatially-modulated ToF and structured light for MPI-free depth estimation." Proceedings of the European Conference on Computer Vision (ECCV) Workshops. 2018.
[6] Amirul Islam, Mohammad Arif Hossain, and Yeong Min Jang, "Interference mitigation technique for time-of-flight (tof) camera,"In 2016 eighth international conference on ubiquitous and future networks (ICUFN), pages 134–139. IEEE, 2016.
[7] A. Bhandari, A. Kadambi, R. Whyte, et al., "Resolving multipath interference in time-of-flight imaging via modulation frequency diversity and sparse regularization,"Opt. Lett. 39(6), 1705–1708 (2014).
[8] A. A. Dorrington, J. P Godbaz, M. J. Cree, et al.,"Separating true range measurements from multipath and scattering interference in commercial range cameras,"In Three-Dimensional Imaging, Interaction, and Measurement, volume 7864, pages 37–46. SPIE, 2011.
[9] S. Fuchs,"Multipath interference compensation in time-of-flight camera images,"In 2010 20th International Conference on Pattern Recognition, pages 3583–3586. IEEE, 2010.

[10] Keel M S, Jin Y G, Kim Y, et al. A VGA Indirect Time-of-Flight CMOS Image Sensor With 4-Tap 7-μm Global-Shutter Pixel and Fixed-Pattern Phase Noise Self-Compensation[J]. IEEE Journal of Solid-State Circuits, 2019, 55(4): 889-897.

[11] Horio M, Feng Y, Kokado T, et al. Resolving multi-path interference in compressive time-of-flight depth imaging with a multi-tap macro-pixel computational CMOS image sensor[J]. Sensors, 2022, 22(7): 2442.

[12] Agresti, Gianluca and Pietro Zanuttigh."Deep Learning for Multi-path Error Removal in ToF Sensors."ECCV Workshops(2018).

[13] Jung H J, Ruhkamp P, Zhai G, et al. On the importance of accurate geometry data for dense 3D vision tasks[C]//Proceedings of the IEEE/CVF Conference on Computer Vision and Pattern Recognition. 2023: 780-791.

[14] Schelling M, Hermosilla P, Ropinski T. Radu: Ray-aligned depth update convolutions for tof data denoising[C]//Proceedings of the IEEE/CVF Conference on Computer Vision and Pattern Recognition. 2022: 671-680.

[15] Duarte M F, Eldar Y C. Structured compressed sensing: From theory to applications[J]. IEEE Transactions on signal processing, 2011, 59(9): 4053-4085.

[16] Xiao Z, Cao S, Zhu L, et al. Channel estimation for movable antenna communication systems: A framework based on compressed sensing[J]. IEEE Transactions on Wireless Communications, 2024.

[17] Li W, Lu W, Liang X, et al. Collaborative Dictionary Learning for Compressed Sensing[J]. IEEE Transactions on Industrial Informatics, 2024.

[18] Candes E J, Tao T. Near-optimal signal recovery from random projections: Universal encoding strategies?[J]. IEEE transactions on information theory, 2006, 52(12): 5406-5425.

[19] S. Li and G. Ge, "Deterministic Sensing Matrices Arising From Near Orthogonal Systems," in IEEE Transactions on Information Theory, vol. 60, no. 4, pp. 2291-2302, April 2014.

[20] F. Tong, L. Li, H. Peng and Y. Yang, "Deterministic Constructions of Compressed Sensing Matrices From Unitary Geometry," in IEEE Transactions on Information Theory, vol. 67, no. 8, pp. 5548-5561, Aug. 2021.

[21] Hu X Y, Eleftheriou E, Arnold D M. Progressive edge-growth Tanner graphs[C]//GLOBECOM'01. IEEE Global Telecommunications Conference (Cat. No. 01CH37270). IEEE, 2001, 2: 995-1001.

[22] Srirutchataboon G, Bajpai A, Wuttisittikulkij L, et al. PEG-like algorithm for LDPC codes[C]//2014 International Electrical Engineering Congress (iEECON). IEEE, 2014: 1-4.

[23] Zhang J, Han G, Fang Y. Deterministic construction of compressed sensing matrices from protograph LDPC codes[J]. IEEE Signal Processing Letters, 2015, 22(11): 1960-1964.

[24] Paredes A L, Conde M H, Loffeld O. Sparsity-aware 3-d tof sensing[J]. IEEE Sensors Journal, 2023, 23(4): 3973-3989.

[25] Tibshirani R. Regression shrinkage and selection via the lasso[J]. Journal of the

Royal Statistical Society Series B: Statistical Methodology, 1996, 58(1): 267-288.

[26] Pati Y C, Rezaiifar R, Krishnaprasad P S. Orthogonal matching pursuit: Recursive function approximation with applications to wavelet decomposition[C]//Proceedings of 27th Asilomar conference on signals, systems and computers. IEEE, 1993: 40-44.

[27] Jiao Y, Jin B, Lu X. Iterative soft/hard thresholding with homotopy continuation for sparse recovery[J]. IEEE Signal Processing Letters, 2017, 24(6): 784-788.

[28] Shahandashti P F, López P, Brea V M, et al. Simultaneous multifrequency demodulation for single-shot multiple-path ToF imaging[J]. IEEE Transactions on Computational Imaging, 2024, 10: 54-68.

[29] Zhang G, Li J, Deng K, et al. Reweighted L1-norm regularized phase retrieval for x-ray differential phase contrast radiograph[J]. Review of Scientific Instruments, 2022, 93(4).

[30] Guo R, Bhandari A. Blind Time-of-Flight Imaging: Sparse Deconvolution on the Continuum with Unknown Kernels[J]. SIAM Journal on Imaging Sciences, 2025, 18(2): 1439-1467.

[31] Bernal-Choban C M, Ladygin V, Granroth G E, et al. Atomistic origin of the entropy of melting from inelastic neutron scattering and machine learned molecular dynamics[J]. Communications Materials, 2024, 5(1): 271.

[32] Ahmed M, Seraj R, Islam S M S. The k-means algorithm: A comprehensive survey and performance evaluation[J]. Electronics, 2020, 9(8): 1295.

[33] Kulkarni A, Mohsenin T. Low overhead architectures for OMP compressive sensing reconstruction algorithm[J]. IEEE Transactions on Circuits and Systems I: Regular Papers, 2017, 64(6): 1468-1480.

[34] Sony Instruments,"SONY IMX570PLR-C", imx570 datasheet, Nov.2020.

[35] Dong G, Zhang Y, Xiong Z. Spatial hierarchy aware residual pyramid network for time-of-flight depth denoising[C]//European conference on computer vision. Cham: Springer International Publishing, 2020: 35-50.

[36] Su S, Heide F, Wetzstein G, et al. Deep end-to-end time-of-flight imaging[C]//Proceedings of the IEEE Conference on Computer Vision and Pattern Recognition. 2018: 6383-6392.

[37] Needell D, Tropp J A. CoSaMP: Iterative signal recovery from incomplete and inaccurate samples[J]. Applied and computational harmonic analysis, 2009, 26(3): 301-321.

[38] Chambolle A, Dossal C H. On the convergence of the iterates of" FISTA"[J]. Journal of Optimization Theory and Applications, 2015, 166(3): 25.